# A topological gauge field in nanomagnets: spin wave excitations over a slowly moving magnetization background

Konstantin Y. Guslienko[1,2*], Gloria R. Aranda[1], and Julian M. Gonzalez[1]

[1]*Dpto. Fisica de Materiales, Universidad del Pais Vasco, 20018 Donostia-San Sebastian, Spain*
[2]*IKERBASQUE, the Basque Foundation for Science, 48011 Bilbao, Spain*

We introduce a topological gauge vector potential which influences spin wave excitations over arbitrary non-uniform, slowly moving magnetization distribution. The time-component of the gauge potential plays a principal role in magnetization dynamics, whereas its spatial components can be often neglected for typical magnetic nanostructures. As an example, we consider spin modes excited in the vortex state magnetic dots. It is shown that the vortex – spin wave interaction can be described as a consequence of the gauge field arising due to non uniform moving vortex magnetization distribution. The coupled equations of motion of the vortex and spin waves are solved within small excitation amplitude approximation. The model yields a giant frequency splitting of the spin wave modes having non-zero overlapping with the vortex mode as well as a finite vortex mass of dynamical origin.

In the recent years a lot of attention was paid to influence of the gauge fields on semi-classical equations of motion of quantum particles such as electrons in solids [1]. It was shown within the local models of itinerant magnetism [2, 3] and *s-d* exchange interaction [4] that a topological gauge field induced by a non-uniform magnetization acts on electron motion as a real external magnetic or electric field [5]. From the other side, spin-polarized electric current can essentially contribute to the equation of motion of magnetization [6] transferring angular momentum. The conception of such gauge field has led to prediction and explanation of various observable magneto-electric effects in patterned nanostructures related to spin angular momentum transfer (spin Hall effect, magnetoresistance, spin pumping, domain wall dynamics etc.), for recent review see Ref. 1 and references therein.

Precise knowledge of the dynamic processes in patterned nanomagnetic materials is important due to their applications in magnetic storage media, magnetic sensors, spintronic devices including arrays of microwave nano-oscillators [7] etc. Understanding dynamic response of nanomagnets to external magnetic field [8-9] or spin-polarized current [10-11] involves rich underlying physics and is also an interesting fundamental issue. Among the different patterned magnetic structures considered so far, flat ferromagnetic particles (dots) with submicron lateral sizes occupy a special place due to their unique, non-uniform vortex ground state [12]. Regular arrays of such dots are being considered as a new high-density non-volatile recording media [13] characterized by two Boolean variables: chirality and polarity of magnetic vortex core [12]. Special interest in the vortex core, small area where magnetization deviates from the dot plane (~10-15 nm in size [14]), is inspired by the possibility of easy and fast switching its magnetization direction [15,16]. The vortex-state dots as well as domain walls in magnetic nanostripes [5, 11] represent the simplest patterned nanosystems with topologically non-trivial magnetization distribution. However, there is still lack of understanding of the spin wave dynamics in such strongly non-uniform ground states.

In this Letter we present a gauge potential approach to description of the spin wave modes excited over arbitrary non-uniform slowly moving background magnetization distribution $\mathbf{M}_\upsilon$. It is shown that a coordinate transformation to the local quantization axis directed along $\mathbf{M}_\upsilon$ allows introducing a topological gauge potential, which leads to an inter-mode interaction even in the linear regime of small amplitude spin excitations.

Let us consider a ferromagnetic body with a time dependent magnetization field $\mathbf{M}(\mathbf{r},t)$ defined in the point $\mathbf{r}$. Assuming conservation of magnetization vector length $|\mathbf{M}| = M_s$ it is convenient to introduce the reduced magnetization $\mathbf{m} = \mathbf{M}/M_s$, $\mathbf{m}^2 = 1$. To describe magnetization dynamics we use the Lagrangian corresponding to the Landau-Lifshitz equation of motion of $\mathbf{m}$:

$$\Lambda = \int d^3\mathbf{r}\,\lambda(\mathbf{r},t), \qquad \lambda = \mathbf{D}(\mathbf{m})\cdot\dot{\mathbf{m}} - w(\mathbf{m},\partial_\alpha\mathbf{m}), \qquad (1)$$

where $\mathbf{D}(\mathbf{m}) = (M_s/\gamma)(1+\mathbf{m}\cdot\mathbf{n})^{-1}[\mathbf{n}\times\mathbf{m}]$ is the vector potential related to the Dirac string in the direction of an arbitrary unit vector $\mathbf{n}$ [17], the dot over symbol means derivative with respect to time ($t$), $w$ is the magnetic energy density $w = A(\partial_\alpha\mathbf{m})^2 + w_m + w_H$ ($\alpha = x, y, z$), $w_m = -M_s\mathbf{m}\cdot\mathbf{H}_m/2$ is the magnetostatic energy density, $w_H = -M_s\mathbf{m}\cdot\mathbf{H}$ is the Zeeman energy density, $A$ is the exchange stiffness, $\mathbf{H}_m$ and $\mathbf{H}$ are the magnetostatic field and external fields.

We distinguish two subsystems in the body: slowly moving magnetization (*e.g.*, a vortex or a domain wall, characteristic time of motion is ~10 ns) + fast magnetization oscillations described as spin waves (SW, with period about of 0.1 ns) and express magnetization as a sum $\mathbf{m} = \mathbf{m}_\upsilon + \mathbf{m}_s$ of the slow magnetization (υ) and spin wave (s) contributions. Recalling the restriction $\mathbf{m}^2 = 1$ the components of $\mathbf{m}_s$ are the simplest in a moving coordinate frame *x'y'z'*, where the axis *Oz'* (quantization axis) is directed along the local direction of $\mathbf{m}_\upsilon$. We assume that the fast s-subsystem can instantly follow slow υ-subsystem magnetization $\mathbf{m}_\upsilon$ (adiabatic approximation), but the spin wave length can be comparable with a characteristic size of $\mathbf{m}_\upsilon$ non-uniformity. Then, we perform a rotation of the initial *xyz* coordinate system to the direction of $\mathbf{m}_\upsilon$. The corresponding 3x3 real rotation matrix $R(\Theta_\upsilon, \Phi_\upsilon) = \exp(i\Phi_\upsilon J_z)\exp(i\Theta_\upsilon J_y)$ is defined by the spherical angles of $\mathbf{m}_\upsilon(\Theta_\upsilon, \Phi_\upsilon)$, and $\mathbf{m}_s$ components in new coordinate system are $\mathbf{m}'_s = R\mathbf{m}_s$, $\mathbf{m}'_\upsilon = (0,0,1)$. Here $J_\alpha$ are the angular momentum components for $J=1$ in the Cartesian basis representation (pure imaginary). To preserve the Lagrangian (1) in the same form after transformation $\mathbf{m} \to \mathbf{m}' = R\mathbf{m}$ we need to introduce to Eq. (1) covariant derivatives $(\partial_\mu - A_\mu)$ instead $\partial_\mu$, where $A_\mu$ is a gauge vector potential (the index µ = 0, 1, 2, 3 denotes the time and space coordinates $x_\mu$ = t, x, y, z, and $\partial_\mu = \partial/\partial x_\mu$). The $A_\mu$ components are transformed as $A_\mu \to A'_\mu = RA_\mu R^{-1} + \partial_\mu R \cdot R^{-1}$. The term $\partial_\mu R \cdot R^{-1}$ has sense of a topological contribution to the vector potential related to the "inertial" moving frame x'y'z'. We denote it as $\hat{A}_\mu = \partial_\mu R \cdot R^{-1}$ and put $A_\mu = 0$ in the laboratory coordinate system xyz. $\hat{A}_\mu$ acts on the SW magnetization and can be represented in the explicit form by the time- and spatial derivatives of the angles $(\Theta_\upsilon, \Phi_\upsilon)$ as

$$\hat{A}_\mu = \begin{pmatrix} 0 & \partial_\mu\Phi_\upsilon & -\cos\Phi_\upsilon\partial_\mu\Theta_\upsilon \\ -\partial_\mu\Phi_\upsilon & 0 & -\sin\Phi_\upsilon\partial_\mu\Theta_\upsilon \\ \cos\Phi_\upsilon\partial_\mu\Theta_\upsilon & \sin\Phi_\upsilon\partial_\mu\Theta_\upsilon & 0 \end{pmatrix}. \qquad (2)$$

The skew-symmetric matrix of the operator $\hat{A}_\mu$ can be represented as $\hat{A}_\mu = A_\mu^\alpha \hat{G}_\alpha$ via the generators $\hat{G}_\alpha$ of SO(3) group of 3-dimensional rotations, which are expressed in the Cartesian basis as $(\hat{G}_\alpha)_{\beta\gamma} = \varepsilon_{\alpha\beta\gamma}$, $\hat{\varepsilon}$ is the unit antisymmetric tensor, α, β, γ= x, y, z. There is a simple equation $\hat{A}_\mu \mathbf{m} = \mathbf{A}_\mu \times \mathbf{m}$ for arbitrary vector $\mathbf{m}$ due to definition of $\hat{A}_\mu$, where $\mathbf{A}_\mu = (\sin\Phi_\upsilon\partial_\mu\Theta_\upsilon, -\cos\Phi_\upsilon\partial_\mu\Theta_\upsilon, -\partial_\mu\Phi_\upsilon)$ is a dual vector. The gauge vector potential $\hat{A}_\mu$ defined by Eq. (2) introduces a "minimal" interaction between the υ- and s-subsystems and can be applied to a wide class of problems related to excitation of the spin waves in the non-uniform magnetization ground state. The first, kinetic, term of the Lagrangian (1) can be written in the frame $x'y'z'$ as $\lambda_{kin} = \mathbf{D}' \cdot (\dot{\mathbf{m}}'_s - \hat{A}_0 \mathbf{m}'_s)$, where the term $\lambda_{int} = -\mathbf{D}' \cdot \hat{A}_0 \mathbf{m}'_s$ corresponds to the dynamic υ – SW interaction. The component $\hat{A}_0$ is always important for spin dynamics, but it was neglected in Refs. 18,19 because the authors considered only static $\mathbf{m}_v$ configurations, where $\hat{A}_\alpha$ appear in the exchange energy terms. Moreover, in the case of body sizes $\gg L_e$ ($L_e = \sqrt{2A}/M_s \sim 10$ nm is the exchange length), $\hat{A}_0$ dominates over other components of $\hat{A}_\mu$ because all the exchange energy terms including spatial derivatives can be neglected, whereas $\hat{A}_0 \propto \dot{\Theta}_v, \dot{\Phi}_v$ determined by the magnetostatic fields is not, in general, small. The Lagrangian (1) then can be re-written in the form $\Lambda = \Lambda_v + \Lambda_{sw} + \Lambda_{int}$, where $\Lambda_{int} = \int d^3 \mathbf{r} \lambda_{int}$ is the interaction term between the slowly moving non-uniform magnetization $\mathbf{m}_v$ described by the Lagrangian $\Lambda_v$ and spin waves, which are described by the Lagrangian density $\lambda_{sw} = \mathbf{D}' \cdot \dot{\mathbf{m}}'_s - w(\mathbf{m}'_s, \partial_\alpha \mathbf{m}'_s)$.

To demonstrate how useful and powerful the approach based on the potential $\hat{A}_\mu$ is, we apply this formalism to the problem of coupled vortex – SW motion in submicron size cylindrical particles (dots) in the vortex state. Magnetic dot with the vortex core possesses two qualitatively different kinds of spin excitation modes: the lowest in frequency gyrotropic mode (typically several 100 MHz) localized near the dot center [9, 20] and high frequency (several GHz) spin waves [9, 10, 21]. The excited mainly outside the vortex core radially and azimuthally symmetric SW modes are described by integers (*n, m*), which indicate number of nodes in the dynamic magnetization along radial (*n*) and azimuthal (*m*) directions. For in-plane magnetic driving field the azimuthal SW and gyrotropic mode ($m = \pm 1$) with non-zero dipolar moments can be only excited [8, 9, 22]. The azimuthal SW with *m*=±1 having the same symmetry as the vortex gyrotropic mode are especially important because they are responsible for the vortex core distortion resulting in the core switching.

The component $A_\mu^z = -\partial_\mu \Phi_v$ of the vector $\mathbf{A}_\mu$ plays crucial role in the dynamics, whereas the other components of $\mathbf{A}_\mu \sim \partial_\mu \Theta_v$, related to the vortex core, can be neglected in the main approximation. The vector potential $\hat{A}_\mu$ leads to essential renormalization of the azimuthal SW spectra and appearance of a dynamical vortex mass induced by the SW-vortex interaction. For the typical dot sizes (the thickness $L \sim 10$ nm, the radius $R \sim 500$ nm) influence of the exchange interaction on dynamics can be neglected. We also assume that $\mathbf{m}$ does not depend on $z$-coordinate along the dot thickness $L$. The vortex magnetization $\mathbf{m}_v(\boldsymbol{\rho},t) = \mathbf{m}_v[\boldsymbol{\rho}, \mathbf{X}(t)]$ can be characterized by its core coordinate $\mathbf{X}=(X, Y)$, velocity, etc., within the collective-variable approach [20]. The unperturbed vortex Lagrangian is

$$\Lambda_v(\mathbf{X}, \dot{\mathbf{X}}) = \frac{1}{2}(\mathbf{G} \times \mathbf{X}) \cdot \dot{\mathbf{X}} - W(\mathbf{X}), \qquad (3)$$

where $W(\mathbf{X})$ is the energy of the vortex shifted from its equilibrium position at $\mathbf{X}=0$ ($\mathbf{H}=0$). $\mathbf{G} = G\hat{\mathbf{z}}$ is the gyrovector, $G = 2\pi p L M_s/\gamma$, $\gamma$ is the gyromagnetic ratio, $\hat{\mathbf{z}}$ is the unit vector perpendicular to the dot plane $xOy$, and $p$ is the vortex core polarization. For submicron dot radii the magnetostatic energy gives the main contribution to $W(\mathbf{X})$.

To derive magnetic vortex and SW dynamics from the Lagrangian $\Lambda$ we use a quadratic approximation in small spin excitation amplitudes and consider linear coupled equations of motion of the system "vortex + spin waves". The vector $\mathbf{m}(\boldsymbol{\rho},t)$ ($\boldsymbol{\rho} = (x, y)$) maps the $xOy$ plane to the surface of unit sphere $\mathbf{m}^2 = 1$. Therefore, we use the angular parameterization for the dot $\mathbf{m}$ components, $m_z = \cos\Theta$, $m_x + im_y = \sin\Theta \exp(i\Phi)$. For the static vortex $\Phi_v(\boldsymbol{\rho},t) = q\varphi + const$, $\Theta_v(\boldsymbol{\rho},t) = \Theta_0(\rho)$, where $\rho, \varphi$ are the polar coordinates. The integer parameter $q$ is the vortex topological charge related to its core (where $\cos\Theta_v \neq 0$). $q=1$ for the simplest vortex. Due to small core size the vortex dynamic magnetization exists mainly outside of the vortex core, in the area where the angle $\Theta_v(\rho) = \pi/2$. We assume that the contribution of the vortex core region to the SW magnetization $\mathbf{m}'_s$ is also small. In the *xyz* frame magnetization $\mathbf{m}(\mathbf{\rho},t)$ is expressed via the angles $\Theta(\mathbf{\rho},t)=\Theta_v(\mathbf{\rho},t)+\vartheta(\mathbf{\rho},t)$, $\Phi(\mathbf{\rho},t)=\Phi_v(\mathbf{\rho},t)+\psi(\mathbf{\rho},t)$, where the variables $\Theta_v(\mathbf{\rho},t)$, $\Phi_v(\mathbf{\rho},t)$ describe the moving vortex, and $\mathbf{m}'_s=(m^s_{x'},m^s_{y'},0)=(\vartheta,\psi,0)$ describes SW excitations. The $\mathbf{m}'_s$-components are used in the form $\vartheta(\mathbf{\rho},t)=a_n(\rho)\cos(m\varphi-\omega t)$, $\psi(\mathbf{\rho},t)=b_n(\rho)\sin(m\varphi-\omega t)$, where $a_n$, $b_n$ are the SW amplitudes, $n=0,1,2…$, $m=0,\pm 1,\pm 2,…$. The dynamic vortex – SW coupling induced by the component $\hat{A}_0$ exists only for the azimuthal modes with $m=\pm 1$, which have the same angular dependence as the vortex gyrotropic mode, and is considered below.

The spin wave Lagrangian density in our particular case is $\lambda_{sw}=(M_s/\gamma)\vartheta\dot\psi+M_s\mathbf{m}_s\cdot\mathbf{H}_m$ and the interaction Lagrangian density is $\lambda_{int}=(M_s/\gamma)\sin\Theta_v\dot\Phi_v\vartheta$. Note that $\lambda_{int}$ is reduced to the form suggested by Slonczewski for infinite films [23]. We get a system of four coupled equations: two integral equations for the spin wave variables $\vartheta$, $\psi$ (neglecting exchange interaction) and two differential ones for the vortex variables $\mathbf{X}=(X,Y)$. The latter are reduced to the Thiele's equation with an extra force $\dot{\mathbf{P}}$ due to the SW field momentum $\mathbf{P}=(M_sL/\gamma)\int d^2\mathbf{\rho}\sin\Theta_0\vartheta\nabla_\mathbf{X}\Phi_v$:

$$\mathbf{G}\times\dot{\mathbf{X}}+\partial_\mathbf{X}W=-\dot{\mathbf{P}}, \tag{4}$$

while the former are

$$\dot\vartheta=-\gamma H^\rho_m, \qquad \dot\psi=\gamma H^z_m-\dot\Phi_v, \tag{4'}$$

where $\mathbf{H}_m(\mathbf{\rho},t)=M_s\int d^2\mathbf{\rho}'\hat{G}(\mathbf{\rho},\mathbf{\rho}')\mathbf{m}_s(\mathbf{\rho}',t)$ is the dynamic magnetostatic field, the kernel $(\hat{G})_{\alpha\beta}=G_{\alpha\beta}(\mathbf{\rho},\mathbf{\rho}')$ is the averaged over thickness magnetostatic tensor [24], $\alpha,\beta=\rho,\varphi,z$.

There is an essential contribution to the SW motion in Eq. (4') due to variable vortex phase $\dot\Phi_v$. The function $\dot\Phi_v$ can be calculated, for instance, by the vortex pole free model [20] to be $\dot\Phi_v=m_0(\rho)[\dot X\sin\varphi-\dot Y\cos\varphi]$, where $m_0(\rho)=(1-\rho^2)/\rho$ is the radial profile of the vortex gyrotropic mode defined in Ref. [21], $\rho$ is in units of *R*. For radial parts of the variables $\vartheta$ and $\psi$ without interaction with the vortex core, we reduce the problem to eigenvalue problem for the integral magnetostatic operator [24] and get a discrete set of magnetostatic eigenfunctions $\boldsymbol{\mu}_{nm}(\boldsymbol{\rho})$ and corresponding SW eigenfrequencies $\omega_{nm}$, which are degenerated with respect to $m$ and are well above the gyrotropic eigenfrequency, $\omega_0$. The SW eigenfrequencies $\omega_{nm}$ are proportional to $(L/R)^{1/2}$ for thin dots, increase with increasing $n$, but decrease with increasing $m$ [12]. The spin eigenfrequencies/eigenfunctions can be found from solution of the inhomogeneous linear integral equation in the main approximation of thin dot $\beta=L/R\ll 1$:

$$\omega^2 a(\rho) - \omega_M^2 \int_0^1 d\rho' \rho' g(\rho, \rho') a(\rho') = \omega_M^2 F(\rho), \qquad (5)$$

where $\omega_M = \gamma 4\pi M_s$, $F(\rho) = \delta_{|m|,1} \int_0^1 d\rho' \rho' g(\rho, \rho') m_0(\rho')$ [24].

Then, expressing the solution of the Thiele's equation for $\mathbf{X}(t)$ (4) via the s-variables and substituting it to Eq. (4') leads to a closed system of equations for the SW amplitudes $a_n(\rho), b_n(\rho)$. Solution of this system yields the perturbed azimuthal SW frequencies $\omega'_{nm} = \omega_{n1} + m\Delta\omega_n/2$ ($m=\pm 1$) and eigenmodes. The frequency splitting is $\Delta\omega_n = p\Im_n^2 \omega_{n1}^2 / 2\omega_M$, where $\Im_n = \int d\rho \rho m_0(\rho) a_n(\rho)$ is the overlapping integral of the vortex gyrotropic mode $m_0(\rho)$ and the unperturbed eigenmode $a_n(\rho)$ obtained from solution of homogeneous Eq. (5), numbered by the radial index $n$ and normalized to unit, and $\omega_{n1}$ are the eigenvalues of homogeneous Eq. (5). $\Im_n$ and $\Delta\omega_n$ are especially big for the main mode $n=0$ reaching 1.6 and ~ 2 GHz for $\beta=0.1$, and they rapidly decrease with $n$ increasing (Fig. 1). The calculated values $\Delta\omega_0 = 1.11$ GHz, $\Delta\omega_1 = 0.99$ GHz for $\beta = 0.048$ with the parameters listed in Ref. [25] are in good agreement with broadband ferromagnetic resonance measurements [22], where $\Delta\omega_0 = 0.87$, $\Delta\omega_1 = 0.80$ GHz were detected for permalloy dots with $L=25$ nm, $R=518$ nm. The SW mode dynamical spatial distributions and accuracy of the expression for $\Delta\omega_n$ were checked numerically [25]. Three resonance peaks of the magnetization response to external in-plane variable field were obtained in the frequency range 0-12 GHz. The first one at 0.48 GHz corresponds to the vortex gyrotropic mode. The other peaks, at 8.98 GHz and 10.44 GHz, are the azimuthal SW ($n=0$; $m=\pm 1$) travelling counter-clockwise and clock-wise, respectively. From these simulations we extracted the spatial distributions of $m_{y'}^s(\boldsymbol{\rho},t)$ for selected times (Fig. 2). The spatial distributions evidence essential difference between the ($m=\pm 1$) azimuthal modes due to strong hybridization with the gyrotropic mode: the low frequency SW mode has radial profile with a peak close to the dot center, whereas the high-frequency mode except the peak near the core has maximum amplitude in the middle of the dot. The presented equation for $\Delta\omega_n$ slightly overestimates the azimuthal SW mode splitting, mainly due to neglecting an intermode dynamical dipolar interaction.

From the other side, expressing the solution of the inhomogeneous SW equation of motion (5) via **X** and substituting it to $\Lambda_{int} = \mathbf{P}\cdot\dot{\mathbf{X}}$ we get the formula $\Lambda_{int} = M_{\alpha\beta}\dot{X}_\alpha \dot{X}_\beta/2$, where $M_{\alpha\beta} = M\delta_{\alpha\beta}$ is the tensor of a vortex mass having the diagonal components $M = -L/(2\gamma^2)\int d\rho\rho \sin\Theta_0 m_0 a$. Solving Eq. (5) we get $a(\rho) = -\sum_n \Im_n a_n(\rho)$ in the limit of the small vortex eigenfrequency $\omega_0 \ll \omega_{n1}$. The vortex mass has the simple form $M = L/(2\gamma^2)\sum_n \Im_n^2$ evidencing its dynamical origin ($M \approx (3/2)L/\gamma^2$ and is about $10^{-20}$ g, being comparable with the transverse domain wall mass $7\cdot 10^{-20}$ g measured in Ref. [26]). All the azimuthal SW with indices ($n$, $m=\pm 1$) contribute to the mass $M$. Therefore, the effective vortex Lagrangian is $\Lambda_v^{eff}(\mathbf{X},\dot{\mathbf{X}}) = M(\dot{\mathbf{X}})^2/2 + (\mathbf{G}\times\mathbf{X})\cdot\dot{\mathbf{X}}/2 - W(\mathbf{X})$. It leads to the gyrotropic frequency $\omega_0' = \omega_0(1-\omega_0 M/4|G|)$, where the correction is relatively small at $\beta \ll 1$, but it explains the deviations from linear behavior $\omega_0(\beta) \propto \beta$ increasing $\beta$ observed in Ref. [20].

The dynamic vortex – SW coupling $\Lambda_{int}$ determined by the time derivatives $\dot{\mathbf{P}}$, $\dot{\Phi}_v$ and expressed via $\Im_n$ exists only for the azimuthal SW modes with $m=\pm 1$ ($n$ is arbitrary). The gauge potential $\hat{A}_0$ generated by the moving vortex core results in the giant frequency splitting $\Delta\omega_n$ of the degenerated azimuthal SW with $m = \pm 1$. The developed approach of the dynamic gauge vortex-SW interaction explains naturally the experimental data [8, 9, 27] on the azimuthal SW frequency splitting. Removing the vortex core from the dot leads to disappearance of the splitting [9, 27] proving importance of the core motion represented by $A_0^z = -\dot{\Phi}_\nu$. The gauge potential $\hat{A}_\mu$ yields the significant spin mode-mode interaction within the linear on spin amplitudes approximation, when the spin eigenmodes are well defined and any non-linear mode-mode interaction can be neglected.

In summary, a general approach to description of the small SW excitations of a non-uniform moving magnetization background was developed within adiabatic approximation and applied to particular case of the spin waves excited in the moving vortex state. The slowly moving magnetization background influences the spin waves via a topological gauge vector potential, which is represented by time and spatial derivatives of the non-uniform magnetization distribution. The vortex – SW interaction results in the giant SW frequency splitting as well as in renormalization of the vortex motion due to appearance of a finite vortex mass. Other dynamic magnetic nanostructures, *e.g.*, moving domain walls in nanostripes (nanorings) can be considered within the approach.

The authors thank A.K. Zvezdin for fruitful discussions. K.G. and G.R.A. acknowledge support by IKERBASQUE (the Basque Science Foundation) and by the Program JAE-doc of the CSIC (Spain), respectively.

**References**

* Corresponding author. E-mail address: sckguslk@ehu.es

**Captions to the Figures**

Figure 1. Dependence of the overlapping integral $\Im_n$ and spin wave azimuthal frequency splitting $\Delta\omega_n$ on the radial mode number $n$. The dot aspect ratio is $\beta = 0.048$, other parameters are as in Ref. 25.

Figure 2. The radial profiles of the main (the radial index $n=0$) azimuthal spin waves in the vortex state with $p= -1$. The eigenmode frequencies are $\omega_{0,+1}/2\pi = 8.98$ GHz and $\omega_{0,-1}/2\pi = 10.44$ GHz for the azimuthal indices $m=+1$ and $m=-1$, respectively. Inset: snapshots of the dynamic $m_{y'}^s$-magnetization component. The dot aspect ratio is $\beta = 0.1$, other parameters are as in Ref. 25.

Fig. 1

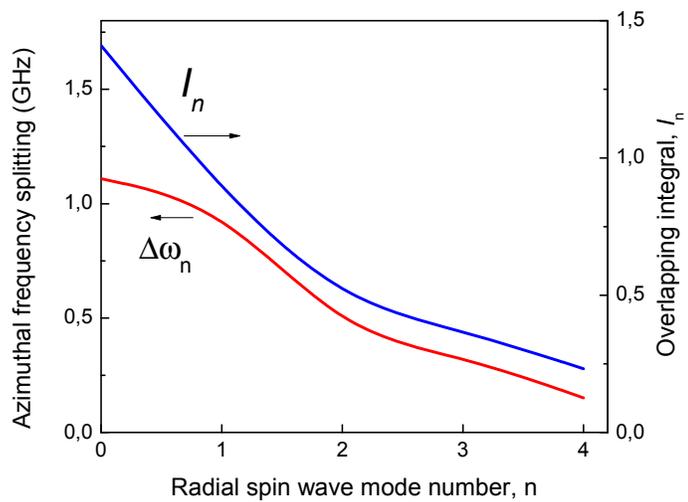

Fig. 2

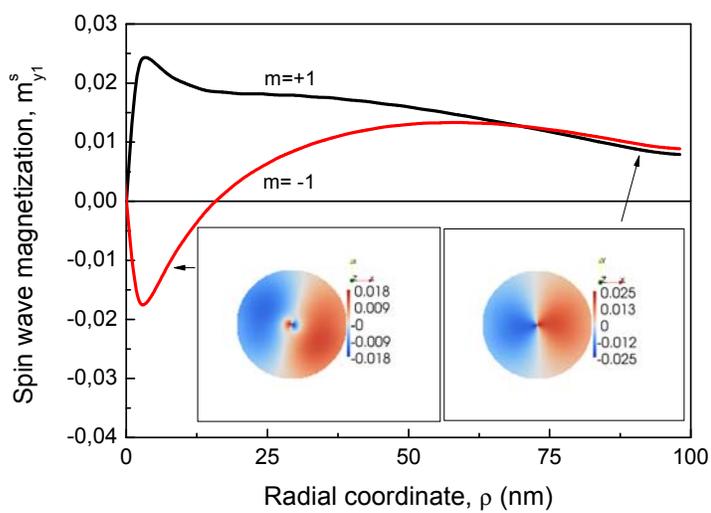